\documentclass[aps,showpacs,preprintnumbers,amsmath,amssymb]{revtex4}

\usepackage{graphicx}
\usepackage{dcolumn}
\usepackage{bm}
\usepackage{hyperref}

\newcommand{\beq}{\begin{equation}}
\newcommand{\eeq}{\end{equation}}
\newcommand{\beqa}{\begin{eqnarray}}
\newcommand{\eeqa}{\end{eqnarray}}
\newcommand{\beqar}{\begin{eqnarray*}}
\newcommand{\eeqar}{\end{eqnarray*}}
\newcommand{\bra}[1]{\mbox{$\left\langle{#1}\right|$}}
\newcommand{\ket}[1]{\mbox{$\left|{#1}\right\rangle$}}

\def\I{{\rm i}}

\newcounter{saveeqn}
\newcommand{\alpheqn}{\setcounter{saveeqn}{\value{equation}}%
\stepcounter{saveeqn}\setcounter{equation}{0}%
\renewcommand{\theequation}{\mbox{\arabic{saveeqn}\alph{equation}}}}
\newcommand{\reseteqn}{\setcounter{equation}{\value{saveeqn}}%
\renewcommand{\theequation}{\arabic{equation}}}
\def\beql{\alpheqn \beqa}
\def\eeql{\eeqa \reseteqn}

\begin{document}
\title{A way to solve BCS-type pairing model}
\author{An Min Wang}
\affiliation{Department of Modern Physics, University of Science
and Technology of China, Hefei, 230026, P.R.China}

\author{Feng Xu}
\affiliation{Department of Modern Physics, University of Science
and Technology of China, Hefei, 230026, P.R.China}




\begin{abstract}
We propose a way to solve BCS-type pairing model by to exactly
solve its spin-analogy in the subspace. The advantages of our
method are to avoid to directly deal with the approximate
procedure and to transfer an exponentially complicated problem to
a polynomial problem. Moreover, it builds the theoretical
foundation of quantum simulation of pairing model and can be used
to check the precision of quantum simulation. It is also helpful
to understand the many-body quantum theory.
\end{abstract}



\maketitle

Recently, the Bardeen, Cooper and Schrieffer (BCS) model for
superconductivity \cite{bcs} has been connected with the problems
in different areas of physics such as superconductivity, nuclear
physics, physics of ultrasmall metallic grains and color
superconductivity in quantum chromodynamics. It is well-known that
BCS-type Hamiltonian is able to be studied by its spin-analogy
form \cite{taylor,solid}. In fact, a famous difficulty of exact
solution of the spin-analogy of pairing model is, in the large-N
limit, the obsession of the exponentially complicated problem. In
addition, the relation the exact solution of spin-analogy of
BCS-type Hamiltonian with the spectrum of BCS-type Hamiltonian
still have to be correctly understood. In this letter, we propose
a way to solve BCS-type pairing model by to exactly solve its
spin-analogy in the subspace. After building the clear relation
between the approximate form of BCS-type pairing model and the
subspace of the spin-analogy of pairing model and finding out the
general form of BCS-type Hamiltonian in this subspace, we are able
to obtain its exact solution in the subspace, that is the
eigenvalues of BCS-type pairing model under the known physical
approximation. The advantages of our method are to avoid to
directly deal with the approximate procedure and to transfer an
exponentially complicated problem to a polynomial problem. It is
also helpful to understand the many-body quantum theory.

Just well-known, a possible way to overcome the exponentially
complicated problem is to use the quantum simulation
\cite{introduceqs}. \textsl{L.-A. Wu} \textsl{et al}.
\cite{simulation} reported a NMR experiment \cite{nmr} scheme
performing a polynomial-time simulation of pairing models. We also
give out a proposal of quantum simulation \cite{Ourqst} which can
obtain the differences of energy levels of pairing model.
Moreover, we have implemented our proposal for the simplest system
of two qubits \cite{Ourqse} in experiment. It must be emphasized
that one has to back (map) to the simulated physical system
\cite{QS}, that is, one would like to know how the spectrum
obtained from quantum simulation corresponds to the physical
spectrum in the simulated physical system, and then one can arrive
at the final aim of quantum simulation. Based on such a cause,
this letter is also be used to shows what is the theoretical
foundation of the quantum simulations including our proposal and
can be used to check the precision of quantum simulation as well
as the physical significance of the results \cite{Our1}.

Let us consider the reduced BCS Hamiltonian \cite{taylor,simulation}:%
\begin{equation}
H_{\mathrm{BCS}}=\sum_{m=1}^{N}\frac{(\varepsilon_{m}-\varepsilon_{F})}%
{2}(n_{m}+n_{-m})-V\sum_{m,l=1}^{N}c_{m}^{\dag}c_{-m}^{\dagger}c_{-l}c_{l}
\label{bcsh}%
\end{equation}
where $n_{\pm m}\equiv c_{\pm m}^{\dagger}c_{\pm m}$ are the
electron number operators, $c_{m}^{\dagger}(c_{m})$ is the
fermionic creation (annihilation) operator. The coupling
coefficient is simplified as a constant $V$
\cite{solidstate,coupleconstant}. Note that the summation indexes
$m=1,2,\cdots,N$ represent all of relevant quantum numbers, and
the electron pairs are labelled by the the quantum number $m$ and
$-m$, according to the Cooper pair situation where the paired
electrons have equal energies but opposite momenta and spins:
$m=(\overrightarrow{k},\uparrow)$ and
$-m=(-\overrightarrow{k},\downarrow)$. Based on Refs.
\cite{taylor,solid},
the pair creation operator is defined by $b_{m}^{\dagger}%
=c_{m}^{\dagger}c_{-m}^{\dagger}$, which generates
$(\overrightarrow {k},\uparrow)$,
$(-\overrightarrow{k},\downarrow)$ electron pair; and the pair
annihilation operator is defined by $b_{m}=c_{m}c_{-m}$, which
destroys $(\overrightarrow{k},\uparrow)$,
$(-\overrightarrow{k},\downarrow)$ electron pair. So one can write
the Hamiltonian (\ref{bcsh}) as \cite{solid}: $
H_{\mathrm{BCS}}=\sum_{m=1}^{N} \xi_{m}(n_{m}+n_{-m})/2-V\sum
_{m,l=1}^{N}b_{m}^{\dagger}b_{l} \label{pairh}%
$, where $\xi_{m}=\varepsilon_{m}-\varepsilon_{F}$ is the free
electron kinetic energy from Fermi surface ($\varepsilon_{F}$ is
the Fermi energy). There are two cases for every pair state $m$:
``occupation" and ``empty", which are denoted respectively by
spin-up state $\chi_{1}=\displaystyle\binom{1}{0}$, spin-down
state $\chi_{0}=\displaystyle\binom{0}{1}$. Introducing the
so-called spin-analogy corresponding of the pair annihilation and
the pair creation operators $b_{m}$ and $b_{m}^\dagger$ as
$b_{m}\Rightarrow\left(
\sigma_{x}^{(m)}-i\sigma_{y}^{(m)}\right)/2 =\sigma_{m}^{-},
b_{m}^{\dagger}\Rightarrow\left(
\sigma_{x}^{(m)}+i\sigma_{y}^{(m)}\right)/2 =\sigma_{m}^{+}$, as
well as the pair number operator $n_{m}+n_{-m}\Rightarrow
1+\sigma_{z}^{(m)}$, one knows that these fermionic pair operators
satisfy the commutation algebra: $sl(2)=\left\{
b_{m},b_{m}^{\dagger},n_{m}+n_{-m}-1\right\}$, \textsl{i.e.}
$sl(2)=\left\{
\sigma_{m}^{-},\sigma_{m}^{+},\sigma_{m}^{z}\right\} $. By use of
the above relations, one can express $H_{\mathrm{BCS}}$
in terms of the spin operators:%
\begin{equation}
H_p^{(N)}=\frac{1}{2}\sum_{m=1}^{N}\epsilon_{m}\sigma_{z}^{(m)}%
-\frac{V}{2}\sum_{m<l=1}^{N}\left(  \sigma_{x}^{(m)}\sigma_{x}^{(l)}%
+\sigma_{y}^{(m)}\sigma_{y}^{(l)}\right)  \label{hspin}%
\end{equation}

where $\epsilon_{m}=\xi_{m}-V$, and the constant term $\displaystyle\frac{1}%
{2}\sum_{m}\epsilon_{m}$ has been ignored \cite{solid}, which
vanishes anyway since we cut off symmetrically above and below
$\varepsilon_{F}$.

From another way, by use of Bogoliubov transformation and mean
field approximation\cite{taylor}, one can obtain the diagonal form
of BSC-type pairing model Hamiltonian
\begin{equation}
\widetilde{H}_{\mathrm{BCS}}=\varepsilon_{s}+\frac{1}{2}\sum_{m=1}^{N}(\xi
_{m}^{2}+\Delta^{2})^{1/2}(\gamma_{m}^{\dagger}\gamma_{m}+\gamma_{-m}%
^{\dagger}\gamma_{-m}) \label{rhbcs}%
\end{equation}
Here $\gamma_{m}^{\dagger}$, $\gamma_{m}$ are the quasiparticle
creation and annihilation operators; $\varepsilon_{s}$ is the
ground state energy of superconducting system. It implies that
there exists such physical approximation that $H_{\rm BCS}$ can be
diagonalized. Note that pairing model Hamiltonian is symmetric and
geminate about $m$ and $-m$, without loss of generality, we are
always able to assume that it is diagonalized as \vskip
-0.2in\beql H_{\rm BCS}^{\rm
diag}&=&\sum_{m=1}^N\sum_{\alpha,\beta=0}\epsilon_{\alpha\beta}
\left(\tilde{n}_m^\alpha\tilde{n}_{-m}^\beta+
\tilde{n}_m^\beta\tilde{n}_{-m}^\alpha\right)\\
&=&\varepsilon_s+\sum_{m=1}^N\frac{E_m}{2}(\tilde{n}_m+\tilde{n}_{-m})
+\sum_{m=1}^N\sum_{\alpha,\beta\neq
0}\epsilon_{\alpha\beta}\left(\tilde{n}_m^\alpha\tilde{n}_{-m}^\beta+
\tilde{n}_m^\beta\tilde{n}_{-m}^\alpha\right)\eeql  where
$\tilde{n}_m^\alpha$ and $\tilde{n}_{-m}^\beta$ are the number
operators of quasiparticle and $\epsilon_{\alpha\beta}$ is a
symmetric tensor. Now, we use the spin-analogy idea to this
diagonalized Hamiltonian and then get \vskip -0.2in\beq
\label{gdiagform} \widetilde{H}_{\rm BCS}^{\rm diag}
=\varepsilon_s+\sum_{m=1}^N\frac{E_m^\prime}{2}(1+\sigma_z^{(m)})+\sum_{m,l=1,m
< l}^N{\epsilon_{m
l}^\prime}(1+\sigma_z^{(m)})(1+\sigma_z^{(l)})\eeq  where we have
used the facts that $[(1+\sigma_z)/2]^n=(1+\sigma_z)/2$ and
$\sigma^{(+)}\sigma^{(-)}=(1+\sigma_z)/2$. Obviously, for the mean
field approximation, $\epsilon_{ml}^\prime\approx 0$ and
$E_m^\prime=(\xi _{m}^{2}+\Delta^{2})^{1/2}$. In fact, since the
success of mean field approximation, one has the reason to believe
that $\epsilon_{ml}^\prime$ almost have no the important effect
for the physical spectrum. Of course, if one can prove that the
diagonalized form of BSC-type pairing model Hamiltonian does not
contain the high order terms about $n_m$ and $n_{-m}$ ($via.$ the
last terms in eq.(\ref{gdiagform})), thus the spectrum in the
subspace 1 of $H_p$ is just the exact solution of BCS-type pairing
model.

Furthermore, in order to understand the relation between the
spectrum of spin-analogy of pairing model and the spectrum of
pairing model itself, we still need to analyse the structure of
the eigenvectors of $H_p$. Note the fact that the spin space can
be divided into the different subspaces which correspond to the
different numbers of spin-up states, that is $S_{\rm
spin}^{(N)}=S_{0}^{(N)}\oplus S_{1}^{(N)}\oplus
S_{2}^{(N)}\oplus\cdots\oplus S_{N}^{(N)}$, where the subspace
$n$, $i.e$ $S_n^{(N)}$, is a subspace with $n$ spin-up states
$\ket{0}$, we can denote the basis of these subspaces as \vskip
-0.2in\beql
S_0^{(N)}&=&\{\ket{2^N}\},\qquad S_N^{(N)}=\{\ket{1}\}\\
S_n^{(N)}&=&\left\{\ket{s^{(N)}_{i_1i_2\cdots
i_n}}=\ket{2^N-\displaystyle\sum_{a=1}^n 2^{(N-i_a)}}, i_a\neq
i_b, i_a=1,2,\cdots,N\right\}\eeql For example, for the subspace 1
of $S^{(N)}$, its basis are
$\ket{s_{i_1}^{(N)}}=\ket{2^N-2^{(N-i_1)}}, (i_1=1,2,\cdots,N)$.
Obviously,
$\left(\ket{2^N-2^{(N-i_1)}}\right)_k=\delta_{k,2^N-2^{(N-i_1)}}$.
It is easy to see that \vskip -0.2in\beq
\bra{s_{i_1}^{(N)}}\widetilde{H}_{\rm BCS}^{\rm
diag}\ket{s_{j_1}^{(N)}}=E_{i_1}^\prime\delta_{i_1j_1}\eeq Then,
let us back to our problem, $via.$ to analyse the structure of
eigenvectors of $H_p$. From
$\left(\sigma_x^{(m)}\sigma_x^{(l)}+\sigma_y^{(m)}\sigma_y^{(l)}\right)
=\left(\sigma_x^{(m)}+\I\sigma_y^{(m)}\right)
\left(\sigma_x^{(l)}-\I\sigma_y^{(l)}\right), (m\neq l)$ in
eq.(\ref{hspin}), it follows that $\ket{1}$ and $\ket{2^N}$ must
be $H_p$'s two eigenvectors respectively corresponding to the
maximum and the minimum eigenvalues. They can be denoted
respectively by $\ket{v_p^1}$ and $\ket{v_p^{2^N}}$. Moreover, for
the arbitrary basis $\ket{s_{i_1\cdots i_n}^{(N)}}$ belongs to
$S_n^{(N)}$, $H_p\ket{s_{i_1\cdots i_n}^{(N)}}$ also belongs to
$S_n^{(N)}$ because that $\sigma^+$ and $\sigma^-$ appear in pairs
or do not appear in the various terms of $H_p$. It implies that
$\bra{s_{i_1\cdots i_{m}}^{(N)}}H_p\ket{s_{i_1\cdots
i_n}^{(N)}}=0, (m\neq n; m,n=1,2,\cdots, N-1)$. Therefore
\begin{equation}
H_{p}^{(N)}=H_{sub0}^{(N)}\oplus H_{sub1}^{(N)}\oplus
H_{sub2}^{(N)}\oplus\cdots\oplus
H_{subN}^{(N)} \label{dirsum1}%
\end{equation}
So we can denote the others eigenvectors of $H_p$ as
$\ket{v_p^{i^{(n)}}}=\sum_{j^{(n)}}b_{i^{(n)}j^{(n)}}\ket{j^{(n)}}\in
S_n^{(N)}$ with the corresponding eigenvalues $E_p^{i^{(n)}}$, and
$i^{(n)},j^{(n)}$ only take the sequence number of spin basis
belonging to the subspace $S_n^{(N)}$, for example
$i^{(1)}(k)=2^N-2^{(N-k)}, (k=1,2,\cdots, N)$. In other words,
$\ket{i^{(n)}},\ket{j^{(n)}}\in S_n^{(N)}$. Actually, above
statement implies that the elements of the diagonalization of
$H_p$ in the subspace 1 are just $E^\prime_{i_1}$. If one only
would like to obtain the spectrum of $H_{\rm BCS}$ in the physical
approximations such as the mean field approximation, it is enough
to find the eigenvalues of the subspace 1 of $H_p^{(N)}$ and does
not need to directly deal with the concrete approximate procedure.

For our propose, not only we need to express the Hamiltonian
$H_p^{(N)}$ as the direct-sum of the submatrices \cite{solid,
simulation}, but also we have to obtain the concrete forms of
$H_{sub1}^{(N)}$. For example, when $N=3$, the basis of subspace
one are $\ket{011},\ket{101}$ and $\ket{110}$. In the spin space,
they are
$(\ket{011})_i=\delta_{i,2^3-2^{(3-1)}}=\delta_{i4},(\ket{101})_i
=\delta_{i,2^3-2^{(3-2)}}=\delta_{i6},(\ket{110})_i
=\delta_{i,2^3-2^{(3-3)}}=\delta_{i7}, (i=1,2,\cdots,8)$. Thus,
denoting $H_p^{(N)}[i,j]$ is an element in the $i$-th row and the
$j$-th column $H_p^{(N)}$, we have \vskip -0.2in\beq
H_{sub1}^{(3)}=\left(
\begin{array}{ccc} H_p^{(3)}[4,4]& H_p^{(3)}[4,6]&H_p^{(3)}[4,7]\\
H_p^{(3)}[6,4]& H_p^{(3)}[6,6]&H_p^{(3)}[6,7]\\
H_p^{(3)}[7,4]& H_p^{(3)}[7,6]&H_p^{(3)}[7,7]\end{array}\right)
=\left( \begin{array}{ccc} \displaystyle\frac{1}{2}(\epsilon_1-\epsilon_2-\epsilon_3)& -V&-V\\
-V& \displaystyle\frac{1}{2}(-\epsilon_1+\epsilon_2-\epsilon_3)&-V\\
-V&
-V&\displaystyle\frac{1}{2}(-\epsilon_1-\epsilon_2+\epsilon_3)\end{array}\right)
\eeq  Likewise, in terms of computer, we can calculate, up to
$N=11$, $H_{sub1}^{(N)}$ has the same structure. That is, all of
the non-diagonal elements are $-V$, the diagonal elements are a
linear combination of all of $\epsilon_i,\ (i,1,2,\cdots,N)$ with
the coefficients the $i$-th being 1/2 and the others are $-1/2$.
That is \vskip -0.2in\beq \label{structurehsub1}
H_{sub1}^{(N)}[i,i]=-\frac{1}{2}\sum_{m=1}^N\epsilon_m+\epsilon_i,\quad
\left. H_{sub1}^{(N)}[i,j]\right|_{i\neq j}=-V  \quad
(i,j=1,2,\cdots,N)\eeq \

Obviously, it is easy to diagonalize above $H_{sub1}^{(N)}$. In
our paper \cite{Our1}, we just used this fact to calculate the
numerical diagonalization of the subspace 1 of $H_p$ up to
$N=100$. Concequentely, the advantage of our method is just to
transfer an exponentially complicated problem to a polynomial
problem. In other words, a task dealing with a $2^{100}$
dimensional matrix is transferred to a task dealing with $100$
dimensional matrix, which can be carried out by the usual
computer.

Now, we would like to prove that $H_{sub1}^{(N)}$ indeed has the
structure in eq.(\ref{structurehsub1}) in terms of mathematical
induction.

We have obtained that the structure of $H_{sub1}^{(3)}$ is true
when $N=3$. Assume that the structure of $H_{sub1}^{(N)}$ is
valid. Consider the case $N+1$, we can rewrite Hamiltonian as
\begin{equation}
H_{p}^{(N+1)}=H_p^{(N)}\otimes I_{2\times 2}
+\frac{1}{2}\epsilon_{N+1}\sigma_{z}^{(N+1)}-
\frac{V}{2}\sum_{m=1}^{N}\sigma_{x}^{(m)}\sigma_{x}^{(N+1)}
-\frac{V}{2}\sum_{m=1}^{N}\sigma_{y}^{(m)}\sigma_{y}^{(N+1)} \label{hspinnew}%
\end{equation}
where $I_{2\times2}$ is a two dimensional identity matrix.

Denoting $p_i^N=2^{N}-2^{N-i},(i=1,2,\cdots, N)$, we have
$(\ket{s_i^{(N)}}_k=\delta_{k,p_i^{N}}$. Obviously, since
$2^{N+1}-2^{N+1-i}=2(2^N-2^{N-i})$, it follows that \vskip
-0.2in\beq \label{position}
\ket{s_{i_1}^{(N+1)}}=\ket{s_{i_1}^{(N)}}\otimes
\left(\begin{array}{c}0\\1\end{array}\right),\quad (i_1\leq N);
\qquad \ket{s_{N+1}^{(N+1)}}=\ket{2^{(N+1)}-1} \eeq It implies
that for a matrix $A$, \vskip -0.2in\beq
A_{p_i^N,p_j^N}=\left(A\otimes
I_{2\times2}\right)_{p_i^{N+1},p_j^{N+1}}, \quad (i\leq N)\eeq
Therefore, the first term $T_1^{(N+1)}=H_p^{(N)}\otimes I_{2\times
2}$ in eq.(\ref{hspinnew}) contributes a same $N$ dimensional
submatrix as $H_{sub1}^{(N)}$.

It is easy to see that only the last component of the $2^N$-th row
and/or the $2^N$-th column of $H_p^{(N)}$ is non zero since
$\left(\sigma_x^{(m)}\sigma_x^{(l)}+\sigma_y^{(m)}\sigma_y^{(l)}\right)
=\left(\sigma_x^{(m)}+\I\sigma_y^{(m)}\right)
\left(\sigma_x^{(l)}-\I\sigma_y^{(l)}\right), (m\neq l)$ in
eq.(\ref{hspin}). Thus, the contribution of the first term
$T_1^{(N+1)}$ to the $(N+1)$-th row and the $(N+1)$-th column of
the submatrix 1 only has a non zero component at the last
(diagonal element) that is the same as the $2^N$-th diagonal
element of $H_p^{(N)}$, the other components are zero.

Obviously the second term
$T_2^{(N+1)}=\epsilon_{N+1}\sigma_{z}^{(N+1)}/2$ in
eq.(\ref{hspinnew}) is a diagonal matrix and for the positions
$p_i^{N+1}, (i=1,2,\cdots, N+1)$, except for $2^{N+1}-1$ gives a
positive $\epsilon_{N+1}/2$, the others are $-\epsilon_{N+1}/2$.
It is easy to see the same conclusion in terms of mathematical
induction when we consider $I_{2\times 2}\otimes
T_2^{(N+1)}=T_2^{(N+2)}$.

To join the contributions to $N\times N$ submatrix in
$H_{sub1}^{(N+1)}$ and the $(N+1)$-th and/or the $(N+1)$-th
column, and add the contributions from the $T_1^{(N+1)}$ and
$T_2^{(N+1)}$, we have \vskip -0.2in\beql
\!\!\!\!\!\!\left(T_1^{(N+1)}+T_2^{(N+1)}\right)_{sub1}\![i,i]\!&=&\!
-\frac{1}{2}\sum_{m=1}^{N+1}\epsilon_m+\epsilon_i,\\
\!\!\!\!\!\! \left.
\left(T_1^{(N+1)}+T_2^{(N+1)}\right)_{sub1}\![i,j]\right|_{i\neq
j,i,j\leq N}\!&=&\!-V,\\
\!\!\!\!\!\!
\left.\left(T_1^{(N+1)}+T_2^{(N+1)}\right)_{sub1}\![i,N+1]\right|_{i\leq
N}
\!&=&\left.\!\left(T_1^{(N+1)}+T_2^{(N+1)}\right)_{sub1}[N+1,j]\right|_{j\leq
N}=0\eeql where we have used that the notation ``$[i,j]$" denotes
the element in the $i$-th row and the $j$ column of a matrix and
subscript ``$i$" denotes the $i$-th component of a vector.

Now, let us consider the last two terms in eq.(\ref{hspinnew}). We
divide three steps to do this.

{\it Step one}: We first prove, in terms of mathematical
induction, that $t_3^{(N)}=\sum_{m=1}^{N}\sigma_{x}^{(m)}$
satisfies: \vskip -0.2in\beq \label{t3sub1} t_3^{(N)}[j,2^N]
=t_3^{(N)}[2^N,j]
=\left(\sum_{i_1=1}^N\ket{s_{i_1}^{(N)}}\right)_j, \quad
\mbox{and}\quad \left(t_3^{(N)}\right)_{sub1}=0 \eeq

For 3 spins we have \vskip -0.2in\beq \label{3spinx}
 \sum_{m=1}^{3}\sigma_{x}^{(m)}=\left(
 \begin{array}{cccccccc}
  0&1&1&0&1&0&0&0\\
  1&0&0&1&0&1&0&0\\
  1&0&0&1&0&0&1&0\\
  0&1&1&0&0&0&0&1\\
  1&0&0&0&0&1&1&0\\
  0&1&0&0&1&0&0&1\\
  0&0&1&0&1&0&0&1\\
  0&0&0&1&0&1&1&0
  \end{array}\right)
  \eeq
obviously eq.(\ref{t3sub1}) is correct. Assume that for $M$ spins,
eq.(\ref{t3sub1}) is also true. Thus, for $M+1$ spins, we have
\vskip -0.2in\beq \label{3spinxnew}
t_3^{(M+1)}=\sum_{m=1}^{M+1}\sigma_{x}^{(m)}=t_3^{(M)}\otimes
I_{2\times 2}+\sigma_{x}^{(M+1)} \eeq From $\left(t_3^{(M)}\otimes
I_{2\times
2}\right)[i,2^{M+1}]=\left[\sum_{\i_1=1}^M\ket{s_{i_1}^{(M)}}\otimes
\left(\begin{array}{c}0\\1\end{array}\right)\right]_i$ and
$\sigma_x^{(M+1)}[i,2^{M+1}]=\delta_{i,2^{M+1}-1}$, it follows
that for $M+1$ spins, the first relation in eq.(\ref{3spinx}) is
valid. Then, notice that $\left(t_3^{(N)}\otimes I_{2\times
2}\right)[p_i^{N+1},p_j^{N+1}]=t_3^{(N)}[p_i^{N},p_j^{N}]=0\
(i,j\leq N)$, $\left(t_3^{(N)}\otimes I_{2\times
2}\right)[p_i^{N+1},2^{M+1}-1]=\left(t_3^{(N)}\otimes I_{2\times
2}\right)[2^{M+1}-1,p_j^{M+1}]=0$, the first term in
eq.(\ref{3spinxnew}) has no contribution to its submatrix 1. While
all of non zero elements of the second term in
eq.(\ref{3spinxnew}) is in its two slanting diagonal lines near
main diagonal line and then has  no contribution to its submatrix
1 either. Therefore, the second relation in eq.(\ref{t3sub1}) for
$M+1$ spins is valid.

{\it Step two}:  Because of eq.(\ref{t3sub1}),
$\left(\sum_{m=1}^{N}\sigma_{x}^{(m)}\sigma_x^{(N+1)}\right)_{sub1}[i,j]=0,
(i,j\leq N)$.

{\it Step three}: First we prove that \vskip -0.2in\beql
\label{3spinx3a}
\left(\sum_{m=1}^{N}\sigma_{x}^{(m)}[2^N,i]\otimes
\sigma_x^{N+1}\right)[1,i]&=&\left(\sum_{i_1=1}^{N+1}\ket{s_{i_1}^{(N+1)}}^T\right)_i\
(i\leq 2^{N+1}-2)\\
\label{3spinx3b}
\left(\sum_{m=1}^{N}\sigma_{x}^{(m)}[2^N,i]\otimes
\sigma_x^{N+1}\right)[1,i]&=&0\quad (i=2^{N+1}-1,2^{N+1}) \eeql
Obviously, for 3 spins, we have \vskip -0.2in\beq
\left(\sum_{m=1}^{3}\sigma_{x}^{(m)}[2^3,i]\otimes
\sigma_x^{3+1}\right)[1,i]=\sum_{i=1}^3(\ket{s_i^{(3)}})^T\otimes\sigma_x^{(3+1)}
=\left(
\begin{array}{cccccccccccccccc}
0&0&0&0&0&0&0&1&0&0&0&1&0&1&0&0\\
0&0&0&0&0&0&1&0&0&0&1&0&1&0&0&0 \end{array}\right) \eeq $i.e$
eqs.(\ref{3spinx3a},\ref{3spinx3b}) are correct.  It must be
emphasized that the right side of
eqs.(\ref{3spinx3a},\ref{3spinx3b}) corresponds to $2^{N+1}-1$
column in the whole matrix. Suppose that for $M$ spins,
eqs.(\ref{3spinx3a},\ref{3spinx3b}) are correct. It is easy to
prove that for $M+1$ spins, this conclusion is correct either
since the first relation in eq.(\ref{t3sub1}) and the first row of
$\sigma_x$ being $(0,1)$. In other word, for arbitrary $N$ spins,
this conclusion is still valid. Therefore, when we extract the
elements of $2^{N+1}-1$ row and the $2^{N+1}-2^{N+1-i}$ columns
($i=1,2,\cdots,N+1$) of $H_p^{(N+1)}$ from the third term in
eq.(\ref{hspinnew}), we obtain a vector whose previous $N$
components are all $-V/2$, the last component is zero. In the same
way, we can prove that when the elements of $2^{N+1}-1$ column and
the $2^{N+1}-2^{N+1-i}$ rows ($i=1,2,\cdots,N+1$) of $H_p^{(N+1)}$
are extracted from the third term in eq.(\ref{hspinnew}), we
obtain a vector whose the previous $N$ components are all $-V/2$
and the last component is zero.

To join the contributions to $N\times N$ submatrix in
$H_{sub1}^{(N+1)}$ and the $(N+1)$-th and/or the $(N+1)$-th column
from
$T_3^{(N+1)}=-V\sum_{m=1}^{N}\sigma_x^{(m)}\sigma_x^{(N+1)}/2$, we
have \vskip -0.2in\beql \left(T_3^{(N+1)}\right)_{sub1}[i,j]=0\ \
(i,j\leq N),\quad , \quad
\left(T_3^{(N+1)}\right)_{sub1}[N+1,N+1]=0 \\
\left(T_3^{(N+1)}\right)_{sub1}[i,N+1]=\left(T_3^{(N+1)}\right)_{sub1}[N+1,i]=-V/2
\ \ (i\leq N) \eeql

By use of the same method, it is able to prove that when we
extract the elements of $2^{N+1}-1$ row and the
$2^{N+1}-2^{N+1-i}$ columns ($i=1,2,\cdots,N+1$) or the
$2^{N+1}-1$ column and the $2^{N+1}-2^{N+1-i}$ rows
($i=1,2,\cdots,N+1$) of $H_p^{(N+1)}$ from the fourth term in
eq.(\ref{hspinnew}), we obtain a vector whose the previous $N$
components are all $-V/2$ and the last component is zero. That is,
the submatrix 1 of
$T_4^{(N+1)}=-V\sum_{m=1}^{N}\sigma_y^{(m)}\sigma_y^{(N+1)}/2$ is
same as one of $T_3^{(N+1)}$.

In the last, we add the all of contributions to the submatrix 1
from every term, our proof in terms of mathematical induction is
finished, $i.e$ the equation (\ref{structurehsub1}) is correct.

As an interesting application of our result, we have calculated
the energy gap in superconductor \cite{Our1} and aimed at the
check the precision of quantum simulation in our proposal of
quantum simulation of pairing model \cite{Ourqst,Ourqse} either.
We believe that there are the other applications in the relevant
problems of pairing model. Moreover, it is possible to extend our
method to the other subspaces as well as the other models of spin
systems.

We thank Zhao Ningbo for his valuable suggestions, and we are
grateful Xiaodong Yang, Xiaosan Ma, Hao You, Niu Wanqing, Zhu
Rengui and Su Xiaoqiang for helpful discussion. This work was
founded by the National Fundamental Research Program of China with
No. 2001CB309310, partially supported by the National Natural
Science Foundation of China under Grant No. 60173047 and the
Natural Science Foundation of Anhui Province.

\end{document}